\documentclass[aps,prb,preprint]{revtex4}
\usepackage[dvips]{graphicx}
\usepackage{amsmath}

\usepackage{endnotes}

\usepackage{amssymb}

\newcommand{\ud}{\textrm{d}}
\providecommand{\abs}[1]{\lvert#1\rvert}
\DeclareMathOperator{\sgn}{sgn}

\newcommand{\eqreff}[1]{Eq.~\eqref{#1}}
\newcommand{\eqrefff}[2]{Eqs.~\eqref{#1} and \eqref{#2}}
\newcommand{\figname}{Fig.}

\begin{document}

\title{The step-harmonic potential}


\author{L. Rizzi}
\affiliation{Dipartimento di Fisica e Matematica, Universit\`a dell'Insubria, Via Valleggio 11, 22100 Como, Italy}
\author{O. F. Piattella}
\affiliation{Dipartimento di Fisica e Matematica, Universit\`a dell'Insubria, Via Valleggio 11, 22100 Como, Italy}
\affiliation{INFN, sezione di Milano, Via Celoria 16, 20133 Milano, Italy}
\author{S. L. Cacciatori}
\affiliation{Dipartimento di Fisica e Matematica, Universit\`a dell'Insubria, Via Valleggio 11, 22100 Como, Italy}
\affiliation{INFN, sezione di Milano, Via Celoria 16, 20133 Milano, Italy}
\author{V. Gorini}
\affiliation{Dipartimento di Fisica e Matematica, Universit\`a dell'Insubria, Via Valleggio 11, 22100 Como, Italy}
\affiliation{INFN, sezione di Milano, Via Celoria 16, 20133 Milano, Italy}

\begin{abstract}
We analyze the behavior of a quantum system described by a one-dimensional asymmetric potential consisting of a step plus a harmonic barrier. We solve the eigenvalue equation by the integral representation method, which allows us to classify the independent solutions as equivalence classes of homotopic paths in the complex plane.
We then consider the propagation of a wave packet reflected by the harmonic barrier and obtain an expression for the interaction time as a function of the peak energy. For high energies we recover the classical half-period limit.
\end{abstract}

\maketitle

\section{Introduction}

The harmonic oscillator plays a central role in physics because it is exactly solvable and provides a simple model for a host of physical phenomena. Besides the simplest case of the one-dimensional free oscillator, modifications of the harmonic oscillator are of interest. A typical variant is to confine the oscillator in a box. This system has been investigated in one dimension\cite{CF1976, ML1983, MC1988, Barton:1990gp, GRD2005} and for arbitrary dimensions.\cite{MAS2007} The purpose of these investigations is to calculate the corrections to the energy levels caused by the presence of infinite barriers at a finite distance. This analysis has also been done for spherically symmetric potentials, such as the hydrogen atom.\cite{Barton:1990gp} In addition, the behavior of a wave packet propagating in a generic power-law one-dimensional potential well has been considered in terms of its ``collapse and revival,'' namely of its scattering over the well and its subsequent reforming.\cite{Robinett2000A, Robinett2000B} The truncated harmonic oscillator has been used to study how the presence of discrete levels in the energy spectrum affects tunneling through such a well.\cite{Chalk1990}

In the present paper we study the bound states and the propagation of a wave packet in a one-dimensional potential consisting of a half-space harmonic oscillator plus a step. In Sec.~\ref{sec:The step-harmonic potential} we solve the Hamiltonian eigenvalue equation using the integral representation method, which allows us to classify the independent solutions as equivalence classes of homotopic paths in the complex plane.
In Sec.~\ref{sec:Eigenfunctions and Energy levels} we calculate the energy of the bound states and compare them with the standard harmonic oscillator and the half-space harmonic oscillator with an infinite barrier. In Sec.~\ref{sec:Scattering and delay} we study the properties of the propagation of a wave packet which, coming from infinite distance, is reflected by the harmonic barrier. Our analysis is based mainly on the investigation of the delay time in the reflection, which can be interpreted as the duration of the interaction with the harmonic barrier. We express the interaction time as a function of the peak energy and study its asymptotic behavior. We show that in the high energy limit the delay approaches the classical value, namely the half period of the harmonic oscillator. Finally, we comment on how this behavior changes in the presence of a stronger or weaker confinement.

The problem that we address has been studied by Mei and Lee\cite{ML1983} to test the adequacy of a perturbation scheme on an exactly solvable model. Our analysis, which has a different purpose, has the advantage of being based on the integral representation method which can be applied to a wider class of problems. In addition, we do not confine ourselves to the study of the bound states, but also investigate the motion of continuous spectrum wave packets.

\section{The step-harmonic potential}\label{sec:The step-harmonic potential}

\begin{figure}[h]
\centering
\includegraphics[scale = 0.5]{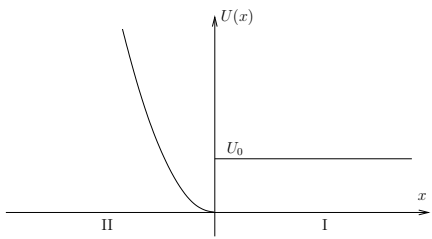}
\caption{The step-harmonic potential.}
\label{Fig:pot}
\end{figure}

Consider a particle subject to the one-dimensional potential:
\begin{equation}\label{eq:pot}
U(x) = \begin{cases}
U_0 & (x \geq 0) \\
\dfrac{1}{2} \kappa x^2 & (x < 0),
\end{cases}
\end{equation}
where $\kappa$ and $U_0$ are real positive constants (see \figname~\ref{Fig:pot}).

The proper and improper eigenfunctions of the Hamiltonian operator are ordinary solutions of the eigenvalue equation outside of the discontinuity of the potential. Such solutions must be continuous together with their first derivatives across the singularity.\cite{Prosperi, griffiths} We will solve the eigenvalue equation for $x > 0$ and $x < 0$.

For $x>0$ the eigenvalue equation is
\begin{equation}\label{SchreqI}
-\dfrac{\hbar^2}{2m} \dfrac{\ud^2 u(x)}{\ud x ^2} + U_0 u(x) = E u(x),
\end{equation}
where $m$ is the particle mass and $E$ is the energy eigenvalue. The general solution of \eqreff{SchreqI} has the form:
\begin{equation}
u(x) = \begin{cases} A e^{ikx} + B e^{-ikx} & (E > U_0) \\
A^\prime e^{kx} + B^\prime e^{-kx} & (0 < E < U_0)
\end{cases},
\end{equation}
where $\hbar k \equiv \sqrt{2m|E - U_0|}$.

We must choose $A^\prime = 0$. Otherwise, for $0 < E < U_0$, $u(x)$ would diverge exponentially for $x \to +\infty$ and therefore it would neither belong to $L^2(\mathbb{R})$ (i.e. the space of the square summable functions over $\mathbb{R}$), 
nor satisfy the eigenpacket condition for improper eigenfunctions.\cite{Prosperi, griffiths}

For $x<0$ the eigenvalue equation is
\begin{equation}\label{SchreqII}
\dfrac{\ud^2 u(y)}{\ud y^2} + (\epsilon - y^2)u(y) = 0,
\end{equation}
where
\begin{equation}
y = \alpha x, \quad \alpha \equiv \sqrt[4]{\dfrac{m\kappa}{\hbar^2}}, \quad \epsilon \equiv \dfrac{2E}{\hbar}\sqrt{\dfrac{m}{\kappa}}, \quad \omega \equiv \sqrt{\dfrac{\kappa}{m}}.
\end{equation}
We set $u(y) = F(y) \exp{(-y^2/2)}$ and obtain from \eqreff{SchreqII} the following equation for $F(y)$
\begin{equation}\label{Hermite}
F^{\prime\prime}(y) - 2y F^{\prime}(y) + (\epsilon - 1)F(y) = 0,
\end{equation}
which is the Hermite equation.\cite{AS1972, Hochstadt1976}

The solutions of \eqreff{Hermite} are entire functions and can be found by the method of integration by series. We prefer to employ the integral representation method.\cite{Hochstadt1976}
We start by looking for solutions of \eqreff{Hermite} that have the form
\begin{equation}\label{prototipo}
F(y) = \!\int_\gamma \ud t\,f(t)e^{-t^2+2ty},
\end{equation}
where $\gamma$ is a path in the complex plane $\mathbb{C}$ and $f$ is a suitable function which is holomorphic in a region which contains the graph of $\gamma$.
We substitute \eqreff{prototipo} into \eqreff{Hermite} and obtain
\begin{equation}\label{eqperconds0}
\int_\gamma\ud t\, [4t^2 + (\epsilon - 1)]f(t)e^{-t^2 + 2ty} - 2\!\int_\gamma\ud t \left(\frac{\ud}{\ud t}e^{2ty}\right)tf(t)e^{-t^2} = 0.
\end{equation}
Equation~\eqref{eqperconds0}, after integration by parts of the second integral, can be written as
\begin{equation}\label{eqperconds}
\left[-2 t f(t) e^{-t^2+2ty}\right]_{\partial\gamma} + \!\int_\gamma\ud t\,[(\epsilon + 1)f(t) + 2tf^\prime(t)]e^{-t^2+2ty} = 0.
\end{equation}
From \eqreff{eqperconds} it follows that \eqreff{prototipo} is a solution of \eqreff{Hermite} if
\begin{equation}\label{conds}
\left[tf(t) e^{-t^2+2ty}\right]_{\partial\gamma} = 0 \quad\text{and}\quad
f(t) = t^{-\tfrac{\epsilon +1}{2}}.
\end{equation}
Therefore, we can write a solution of \eqreff{Hermite} in the form
\begin{equation}\label{gensol}
F^{(\gamma)}(y) = \!\int_\gamma\ud t\, t^{-\tfrac{\epsilon + 1}{2}} e^{-t^2+2ty},
\end{equation}
where $\gamma$ must be chosen according to the first condition in \eqreff{conds} and such that the integral in \eqreff{gensol} is well defined. The classification of the appropriate $\gamma$'s allows us to classify all solutions of \eqreff{Hermite}.

The integrand in \eqreff{gensol} is singular at $t = 0$. For $\epsilon = 2n + 1$ ($n = 0, 1, \dots$), the point $t=0$ is a pole of order $n + 1$, otherwise it is a branch point.

In the following we distinguish two classes of paths, which correspond to two linearly independent solutions of \eqreff{Hermite}.

\subsection{The case $\epsilon = 2n+1$}

\begin{figure}
\centering
\includegraphics[scale=0.8]{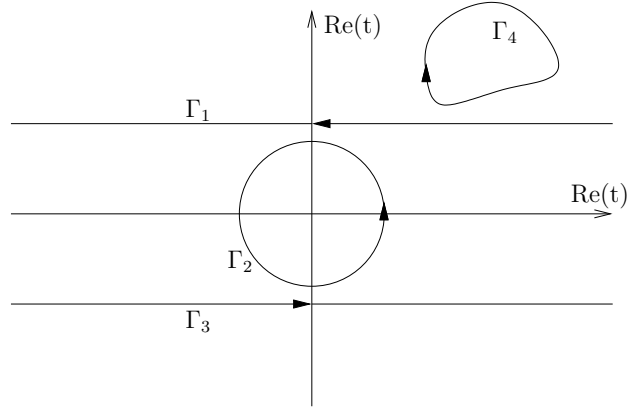}
\caption{Possible paths for $\epsilon = 2n+1$ ($n=0,1,\dots$).}\label{cammini}
\end{figure}

In this case we can rewrite \eqreff{prototipo} as
\begin{equation}\label{simple}
F_n^{(\gamma)}(y) = \!\int_\gamma \ud t\,\dfrac{e^{-t^2+2ty}}{t^{n+1}},
\end{equation}
where the integrand is holomorphic on $\mathbb{C}$ but the origin.
Possible choices of $\gamma$ for which the contour condition in \eqreff{conds} holds are shown in \figname~\ref{cammini}; $\Gamma_1$ and $\Gamma_3$ have a real part that goes to infinity, $\Gamma_2$ is a closed path circling the origin, and $\Gamma_4$ is a closed path that does not contain the origin. By virtue of Cauchy's theorem, $F_n^{(4)} = 0$, and because the paths can be deformed so that $\Gamma_1 + \Gamma_3 = \Gamma_2$, the other three solutions satisfy the relation
$F_n^{(1)} + F_n^{(3)} = F_n^{(2)}$,
where $F_n^{(j)}$ is the solution corresponding to the path $\Gamma_j$ ($j = 1,2,3,4$). Then we have, as expected, two linearly independent solutions for \eqreff{Hermite}.

As an exercise, we show that the solution corresponding to $\Gamma_2$, namely
\begin{equation}\label{tobeHermite}
F^{(2)}_n(y) = \!\oint\ud t\,\dfrac{e^{-t^2+2ty}}{t^{n+1}},
\end{equation}
corresponds to the Hermite polynomial of order $n$. By completing the square in the integrand of \eqreff{tobeHermite}, we find
\begin{equation}\label{tobeHermite1}
F^{(2)}_n(y) = e^{y^2}\!\oint \ud t\,\dfrac{e^{-(t-y)^2}}{t^{n+1}}.
\end{equation}
We take advantage of Cauchy's formula and rewrite \eqreff{tobeHermite1} as
\begin{equation}
F^{(2)}_n(y)= \dfrac{2\pi i}{n!}(-1)^n\,e^{y^2}\dfrac{\ud^n}{\ud y^n}(e^{-y^2}) = \dfrac{2\pi i}{n!}H_n(y),
\end{equation}
where $H_n(y)$ is the Hermite polynomial of order $n$.\cite{Hochstadt1976}

\subsection{The case $\epsilon\neq2n+1$}

\begin{figure}
\centering
\includegraphics[scale=0.8]{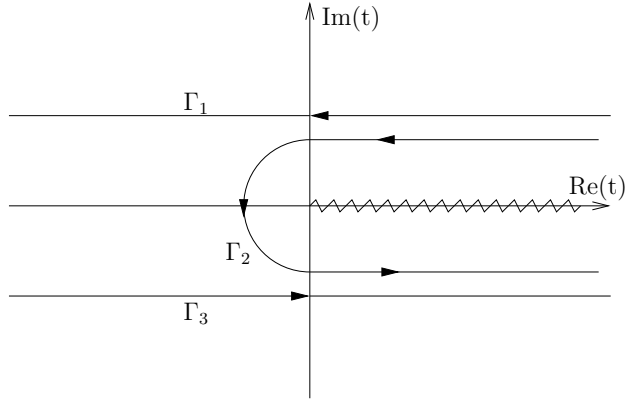}
\caption{Possible paths for $\epsilon \neq 2n + 1$. Note that in this case $\Gamma_2$ cannot be closed at infinity.}\label{cammini2}
\end{figure}

In the generic case $\epsilon \in \mathbb{R}$, $\epsilon\neq2n+1$, we rewrite \eqreff{gensol} as
\begin{equation}\label{simple2}
F^{(\gamma)}_\epsilon(y) = \!\int_\gamma \ud t\,\dfrac{e^{-t^2 +2ty}}{t^\beta},
\end{equation}
where $\beta \equiv (\epsilon + 1)/2$.
If $\beta$ is not a positive integer, $t = 0$ is a branch point for the multivalued function $t^\beta$. In this case we must cut the complex plane, for example along the positive real axis. In the latter case, the classes of possible paths are depicted in \figname~\ref{cammini2}.

In the following we show that the solutions corresponding to $\Gamma_1$ and $\Gamma_3$ ($F^{(1)}_\epsilon$ and $F^{(3)}_\epsilon$) diverge as $e^{y^2}$ for $y = \pm \infty$ and therefore the corresponding eigenfunction $u(y)$ cannot be either proper or improper. We are thus left with the solution corresponding to $\Gamma_2$ [$F^{(2)}_\epsilon$] which again diverges as $e^{y^2}$ for $y\rightarrow+\infty$. It also diverges for $y\rightarrow-\infty$, but the corresponding $u(y)$ and its derivatives vanish more rapidly than any polynomial thanks to the presence of the $\exp(-y^2/2)$ factor.

According to standard results in the theory of integrals depending on a parameter, it is easy to show that $F^{(\gamma)}_\epsilon(y)$ is an entire function. Furthermore, the derivatives of $F^{(\gamma)}_\epsilon$ are obtained by differentiating with respect to $y$ under the integral sign of \eqreff{simple2}. Thus we obtain the relation
\begin{equation}\label{derivativesrelation}
\dfrac{\ud^m F^{(j)}_\epsilon}{\ud y ^m} = 2^m F^{(j)}_{\epsilon-2m}.
\end{equation}
We next address the asymptotic behavior of two independent solutions, for example, $F^{(1)}_\epsilon$ and $F^{(2)}_\epsilon$.

{\it The $\Gamma_1$ solution}. We rewrite \eqreff{simple2} for the path $\Gamma_1$ by introducing the variable $z = t - y$:
\begin{equation}\label{gamma1sol}
F^{(1)}_\epsilon(y) = e^{y^2}\int_{\Gamma_1}\ud z\, \dfrac{e^{-z^2}}{(y+z)^\beta}.
\end{equation}
The branch point is now $z = -y$ and the cut is shifted as well.
We extract $\abs{y}$ from the integral in \eqreff{gamma1sol}\cite{foot1}:
\begin{equation}
F^{(1)}_\epsilon(y) = \dfrac{e^{y^2}}{\abs{y^\beta}}G_1(y),
\end{equation}
where
\begin{equation}
G_1(y) \equiv \!\int_{\Gamma_1}\ud z\, \dfrac{e^{-z^2}}{[\sgn(y)+\frac{z}{\abs{y}}]^\beta}.
\end{equation}
An elementary calculation shows that
\begin{equation}
\lim_{y\rightarrow\pm\infty}G_1(y) = \lim_{y\rightarrow\pm\infty}\int_{\Gamma_1}\ud z\, \dfrac{e^{-z^2}}{[\sgn(y)+\frac{z}{\abs{y}}]^\beta} = -\dfrac{\sqrt{\pi}}{\sgn(y)^\beta},
\end{equation}
where we have taken the limit under the integral sign by virtue of the dominated convergence theorem\cite{C1967} (see \appendixname~\ref{appendixB}). The asymptotic behavior of $F^{(1)}_\epsilon(y)$ for $y\rightarrow\pm\infty$ is therefore
\begin{equation}
F^{(1)}_\epsilon(y)\sim-\sqrt{\pi}\dfrac{e^{y^2}}{y^\beta}.
\end{equation}
The corresponding eigenfunction $u^{(1)}(y) = F^{(1)}_\epsilon(y)\exp\left(-y^2/2\right)$ cannot be either proper or improper.

{\it The $\Gamma_2$ solution}. The solution corresponding to $\Gamma_2$ [$F^{(2)}_\epsilon(y)$] has the form
\begin{equation}\label{secondindsol}
F^{(2)}_\epsilon(y) = \!\int_{\Gamma_2} \ud t\,\dfrac{e^{-t^2 +2ty}}{t^\beta},
\end{equation}
where, as shown in \figname~\ref{cammini2}, $\Gamma_2$ circles around the branch point in an anti-clockwise sense. This solution has different behavior for $y\rightarrow + \infty$ and $y\rightarrow - \infty$.

\paragraph*{Asymptotic behavior for $y\rightarrow+\infty$.}

By virtue of Cauchy's theorem we can deform $\Gamma_2$ to split the integral in \eqreff{secondindsol} into a sum of two integrals over the paths $\Gamma_1$ and $\Gamma_3$:
\begin{equation}
F^{(2)}_\epsilon(y) = \!\int_{\Gamma_1\cup\Gamma_3}\ud t\, \dfrac{e^{-t^2 +2ty}}{t^\beta}.
\end{equation}
We again introduce $z = t - y$ and extract $y^\beta$ from the integral. We obtain
\begin{equation}
F^{(2)}_\epsilon(y) = \dfrac{e^{y^2}}{y^\beta}G_2(y),
\end{equation}
where
\begin{equation}
G_2(y) \equiv \!\int_{\Gamma_1}\ud z\,\dfrac{e^{-z^2}}{(1+\frac{z}{y})^\beta} + \!\int_{\Gamma_3}\ud z\,\dfrac{e^{-z^2}}{(1+\frac{z}{y})^\beta}.
\end{equation}
Thanks to the dominated convergence theorem we find
\begin{equation}
\lim_{y\rightarrow+\infty}G_2(y) = -2ie^{-i\pi\beta}\sqrt{\pi}\sin(\pi\beta).
\end{equation}
The asymptotic behavior of $F^{(2)}_\epsilon(y)$ for $y\rightarrow+\infty$ is thus
\begin{equation}\label{f2asymptyinf}
F^{(2)}_\epsilon(y)\sim-2ie^{-i\pi\beta}\sqrt{\pi}\sin(\pi\beta)\dfrac{e^{y^2}}{y^\beta}.
\end{equation}
Note that, if $\epsilon = 2n + 1$, \eqreff{f2asymptyinf} is incorrect because $G_2(y) \to 0$ for $y\rightarrow+\infty$. We already know that, in this case, $F^{(2)}(y)$ is as a polynomial of degree $n$.

\paragraph*{Asymptotic behavior for $y\rightarrow-\infty$.}
\begin{figure}
\centering
\includegraphics[scale = 0.8]{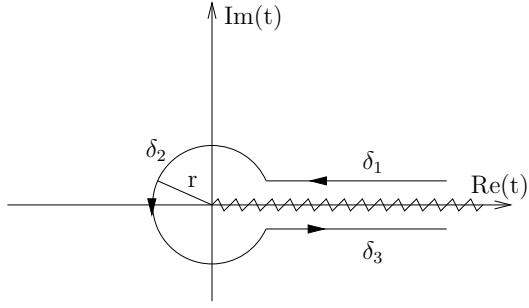}
\caption{A standard trick in contour integration.}\label{split}
\end{figure}
By taking advantage of Cauchy's theorem, we deform and split $\Gamma_2$ in the 3 sub-paths shown in \figname~\ref{split}. We choose for simplicity $r = 1$; from \eqreff{secondindsol} we obtain
\begin{equation}\label{eqn:bordello2}
F^{(2)}_\epsilon(y) = I_\beta(y) - 2i e^{-i\pi\beta}\sin(\pi\beta)\!\int_1^{\infty} \ud t\, \dfrac{e^{-t^2 +2ty}}{t^\beta},
\end{equation}
where
\begin{equation}
I_\beta(y) \equiv i\!\int_0^{2\pi} \ud \theta\, e^{i(1-\beta)\theta} e^{-\cos(2\theta)-i\sin(2\theta)+2y\cos\theta + 2iy\sin\theta}.
\end{equation}
For $y < 0$ and $t > 0$ the following inequality holds
\begin{equation}
t^{-\beta} e^{-t^2+2ty} < t^{-\beta}e^{-t^2},
\end{equation}
and, moreover, $t^{-\beta}\exp(-t^2)$ is integrable in $[1,+\infty)$. Therefore, the integral on the right-hand side of \eqreff{eqn:bordello2} vanishes for $y\rightarrow-\infty$ by virtue of the dominated convergence theorem.

In contrast, for $I_\beta(y)$ we have
\begin{equation}
\abs{I_{\beta}(y)}\leq \!\int_0^{2\pi}\ud\theta\,\Big\lvert e^{i(1-\beta)\theta} e^{-\cos(2\theta)-i\sin(2\theta)+2y\cos\theta + 2iy\sin\theta}\Big\rvert = \!\int_0^{2\pi}\ud\theta\, e^{-\cos(2\theta)}e^{2y\cos\theta},
\end{equation}
so that $\abs{I_\beta(y)} \leq 2\pi e^{2\abs{y}-1}$.
Therefore, for $y\rightarrow-\infty$, the absolute value of $F^{(2)}_\epsilon(y)$ is dominated by $2\pi e^{2\abs{y}-1}$, and
\begin{equation}
u_{\rm{II}}(y) = F^{(2)}_\epsilon(y) e^{-y^2/2}
\end{equation}
is rapidly decreasing. Hence it is square summable on the positive real axis.
Thus, it follows that for $\epsilon\neq 2n+1$ the full-space harmonic oscillator does not admit proper or improper eigenfunctions. More general theorems\cite{BC:1980} allow one to obtain our results indirectly, for example, by studying the asymptotic behavior of the power series expansion of the solutions of \eqreff{SchreqII}. Here we have adopted a more direct approach.

\section{Eigenfunctions and Energy levels}\label{sec:Eigenfunctions and Energy levels}

From the results of Sec.~\ref{sec:The step-harmonic potential} we can write the energy eigenfunctions as
\begin{align}
& E<U_0 &
u(x) = & \begin{cases}
A F_\epsilon (\alpha x)e^{-\tfrac{\alpha^2 x^2}{2}} & (x<0) \\
B e^{-kx} & (x>0)
\end{cases} \label{E<U}\\
& E>U_0 &
u(x) = & \begin{cases}
C F_\epsilon (\alpha x)e^{-\tfrac{\alpha^2 x^2}{2}} & (x<0) \\
D e^{ikx} + E e^{-ikx} & (x>0),
\end{cases} \label{E>U}
\end{align}
where we have dropped the superscript from $F^{(2)}_\epsilon$. The integration constants $A, \dots, E$ must be chosen such that $u(x)$ and its first derivative are continuous at $x = 0$ (the junction conditions).

\subsection{The case $E < U_0$}

If the energy is smaller than the step height $U_0$, the junction conditions imply that
\begin{subequations}
\begin{align}
B - F_\epsilon(0) A &= 0 \\
k B + \alpha F_\epsilon^\prime(0) A &= 0.
\end{align}
\end{subequations}
The condition for the existence of a nontrivial solution is the vanishing of the system determinant. We define $J(\beta) \equiv F_\epsilon(0)$ [see \eqreff{Jbeta}]. In \appendixname~\ref{appendixA} we obtain the following expression for $J(\beta)$ [see \eqreff{J}]:
\begin{equation}\label{Jbeta_first}
J(\beta) = \dfrac{\sin(\pi\beta)}{ie^{i\pi\beta}}\Gamma\left(\dfrac{1-\beta}{2}\right).
\end{equation}
The recurrence relation for the derivatives of $F_\epsilon$, \eqreff{derivativesrelation}, can be rewritten in terms of $J(\beta)$ as
\begin{equation}\label{thiswhat}
F_\epsilon^\prime(0) = 2 J (\beta - 1).
\end{equation}
Equation~\eqref{thiswhat} implies that the junction conditions can be rewritten as
\begin{equation}\label{junccond}
-2\alpha J(\beta -1) = k J(\beta),
\end{equation}
or, equivalently, as
\begin{equation}\label{eqlivelli1}
\dfrac{\Gamma\left(1-\dfrac{\beta}{2}\right)}{\Gamma\left(\dfrac{1-\beta}{2}\right)} = -\sqrt{\dfrac{\beta_0-\beta}{2}},
\end{equation}
where $\beta_0 = U_0/(\hbar\omega) + 1/2$. By using the relation\cite{AS1972}
\begin{equation}\label{gammaformula}
\Gamma(z)\Gamma(1-z) = \dfrac{\pi}{\sin(\pi z)},
\end{equation}
we can rewrite \eqreff{eqlivelli1} as
\begin{equation}\label{eqlivelli2}
\dfrac{\Gamma\left(\dfrac{\beta+1}{2}\right)}{\Gamma\left(\dfrac{\beta}{2}\right)}\cot\left(\dfrac{\pi}{2}\beta\right) = -\sqrt{\dfrac{\beta_0-\beta}{2}},
\end{equation}
For a given value of $\beta_0$ Eq.~\eqref{eqlivelli2} is an implicit relation determining the energy levels. The advantage of \eqreff{eqlivelli2} is that the singular behavior is contained in the cotangent function.

\begin{figure}
\centering
\includegraphics[scale=0.8]{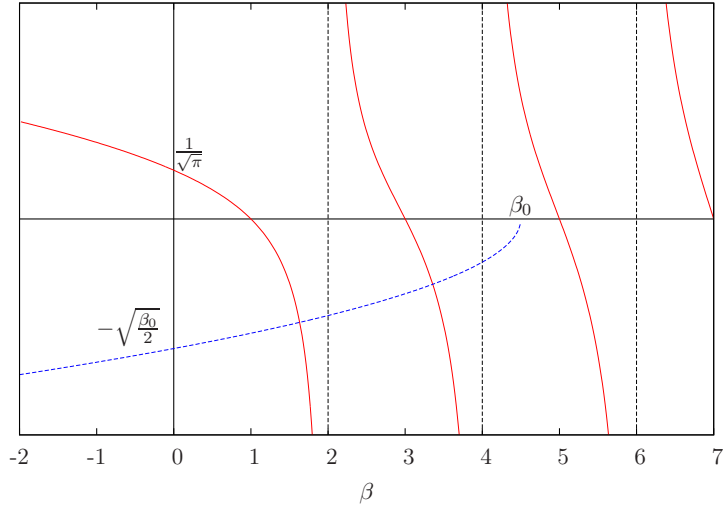}
\caption{The solid and the dashed lines represent respectively the left- and the right-hand side of \eqreff{eqlivelli2}. The intersections determine the energy levels. Here $\beta_0 = 4.5$.}\label{energylevels}
\end{figure}

For $0\leq U_0 < \hbar\omega / 2$ ($1/2\leq \beta_0 < 1$) the step is too small to allow for the existence of discrete energy levels. The first level (the ground state) appears for $\beta_0 = 1$ at the value $E_0 = \hbar\omega/2$, which is the ground state of the full-space harmonic oscillator. For $1 \leq \beta_0 < 3$ there is only one level, whose energy grows with increasing $\beta_0$ starting from its minimum value $\hbar\omega/2$. When $\beta_0$ crosses the value $3$ a second level appears at the energy $E_1 = 5\hbar\omega /2$. By further increasing $\beta_0$ (and hence $U_0$) the subsequent levels appear as $U_0$ crosses the values $E_k = \hbar\omega(2k+1/2)$ ($k\in\mathbb{N}$), corresponding to the $(k+1)$th even level of the oscillator. Thus, for a fixed value of $\beta_0$ such that $2k + 1< \beta_0 < 2k + 3$ there are exactly $k + 1$ bound states with energies $E_n$ ($n = 0,1,\dots, k$) satisfying the inequalities $\hbar\omega(2n+1/2)< E_n < \hbar\omega(2n+3/2)$. An example with $k=1$ is shown in \figname~\ref{energylevels}. Each $E_n$ is a monotonically increasing function of $U_0$ which asymptotically approaches the value $\hbar\omega(2n+3/2)$ (the $(n+1)$th odd state of the oscillator) as $U_0\to\infty$.

The (unnormalized) eigenfunctions corresponding to the eigenvalue $E_n$ are
\begin{equation}\label{proper_eigenfunctions}
u_n(x) = \begin{cases}
F_{\epsilon_n} (\alpha x)e^{-\tfrac{\alpha^2 x^2}{2}} & (x<0) \\
J(\beta_n) e^{-k_n x} & (x\geq0),
\end{cases}
\end{equation}
where the $\beta_n$ are the solutions of \eqreff{eqlivelli2}; $\epsilon_n = 2\beta_n -1$, $\hbar k_n = \sqrt{2m(U_0-E_n)}$, and $E_n = \hbar\omega\epsilon_n/2$.

\subsection{The case $E > U_0$}

The junction conditions on the eigenfunctions of \eqreff{E>U} are
\begin{subequations}
\begin{align}
D+E - CF_\epsilon(0) &= 0 \\
i k (D - E) - C\alpha F_\epsilon^\prime(0) & = 0,
\end{align}
\end{subequations}
implying the normalized (with respect to $k$) improper eigenfunctions are given by
\begin{equation}\label{improper_eigenfunctions}
u_\epsilon(x) = \dfrac{1}{\sqrt{2\pi}}\begin{cases}
\Pi(\beta)F_\epsilon (\alpha x)e^{-\tfrac{\alpha^2 x^2}{2}} & (x<0) \\
e^{-ikx} + \zeta(\beta) e^{ikx} & (x \geq 0),
\end{cases}
\end{equation}
where, as usual, $2\beta \equiv \epsilon + 1$, $\hbar k \equiv \sqrt{2m(E - U_0)}$, $2E \equiv \hbar\omega\epsilon$, and
\begin{align}
\Pi(\beta) &\equiv 2 \left[ J(\beta)+i\sqrt{\tfrac{2}{\beta-\beta_0}}J(\beta-1)\right]^{-1},\\
\label{zeta}
\zeta(\beta) &\equiv \dfrac{J(\beta)-i\sqrt{\tfrac{2}{\beta-\beta_0}}J(\beta-1)}{J(\beta)+i\sqrt{\tfrac{2}{\beta-\beta_0}}J(\beta-1)} = \dfrac{\Gamma\left(\dfrac{1-\beta}{2}\right)-i\sqrt{\tfrac{2}{\beta-\beta_0}}\Gamma\left(\dfrac{2-\beta}{2}\right)}{\Gamma\left(\dfrac{1-\beta}{2}\right)+i\sqrt{\tfrac{2}{\beta-\beta_0}}\Gamma\left(\dfrac{2-\beta}{2}\right)},
\end{align}
where we have used \eqreff{Jbeta}. Note that $\abs{\zeta(\beta)} = 1$. As expected, the continuous part of the spectrum ($E > U_0$) is simple.

\section{Reflection and delay}\label{sec:Scattering and delay}

To study reflection phenomenon, we consider the following superposition of continuous states:
\begin{equation}
\psi(x,t) = \!\int_0^{\infty}\ud k\, c(k) u_{\epsilon(k)}(x) e^{-\tfrac{i}{\hbar}E(k) t}.
\end{equation}
From \eqreff{improper_eigenfunctions} we have
\begin{equation}
\psi(x,t) = \frac{1}{\sqrt{2\pi}}
\begin{cases}
\int_{0}^{\infty}\ud k\, c(k) \Pi(\beta(k)) F_\epsilon (\alpha x) e ^{-\tfrac{\alpha^2 x^2}{2}-\tfrac{i}{\hbar}E(k) t} & (x < 0) \\
\int_{0}^{\infty}\ud k\, c(k) \left[\zeta(\beta(k)) e ^{ikx} + e ^{-ikx}\right]e^{-i \tfrac{E(k)}{\hbar} t} = \psi_{\text{ref}}+\psi_{\text{in}} & (x > 0)
\end{cases}.
\end{equation}
We write $\psi_{\text{in}}$ and $\psi_{\text{ref}}$ in the form:
\begin{align}
\psi_{\text{in}}(x,t) & = \frac{1}{\sqrt{2\pi}}\!\int_{0}^{+\infty}\ud k\, \abs{c(k)} e^{-i[kx+\Omega(k) t - \gamma(k)]}, \\
\psi_{\text{ref}}(x,t) & = \frac{1}{\sqrt{2\pi}}\!\int_{0}^{+\infty}\ud k\, \abs{c(k)} e^{i[kx-\Omega(k) t + \delta(k)+ \gamma(k)]},
\end{align}
where we have defined
\begin{equation}
e^{i\delta(k)} \equiv \zeta(\beta(k)) \quad \text{and} \quad
\Omega(k) \equiv \dfrac{E(k)}{\hbar} = \dfrac{U_0}{\hbar} + \dfrac{\hbar k^2}{2m}.
\end{equation}
If $c(k)$ is sufficiently regular and non-vanishing only in a small neighborhood of $\tilde{k}$, then $\psi_{\text{in}}$ and $\psi_{\text{ref}}$ represent wave packets that move according to the equations of motion\cite{Prosperi, griffiths}
\begin{equation}
x_{\text{in}} = -\left.\dfrac{\ud \Omega}{\ud k}\right|_{k=\tilde{k}}t +\left.\dfrac{\ud \gamma}{\ud k}\right|_{k=\tilde{k}} = - \dfrac{\hbar \tilde{k}}{m} (t-t_0)= -\dfrac{\tilde{p}}{m} (t-t_0),
\end{equation}
for the ``incoming'' wave packet, and
\begin{equation}
x_{\text{ref}} = \left.\dfrac{\ud \Omega}{\ud k}\right|_{k=\tilde{k}} t -\left.\dfrac{\ud \gamma}{\ud k}\right|_{k=\tilde{k}} - \left.\dfrac{\ud \delta}{\ud k}\right|_{k=\tilde{k}}= \dfrac{\tilde{p}}{m} \left[(t-t_0)-\dfrac{m}{\tilde{p}}\left.\dfrac{\ud \delta}{\ud k}\right|_{k=\tilde{k}}\right],
\end{equation}
for the reflected ``outgoing'' one.

The solution represents a particle of well defined momentum $\tilde{p}=\hbar\tilde{k}$ which approaches the origin from the right, interacts with the harmonic potential (at $t=t_0$), and is totally reflected. The phase shift results in a delay in the time the wave packet bounces back, which is caused by the interaction with the confining harmonic barrier. Because the phase shift $\delta$ depends only on $k$ through $\beta$, we can write the delay as
\begin{equation}\label{delay}
\tau(\tilde{\beta}) = \dfrac{1}{\omega}\left.\dfrac{\ud \delta}{\ud \beta}\right|_{\beta=\tilde{\beta}},
\end{equation}
where $\tilde{\beta} = \beta(\tilde{k})$.

\begin{figure}
\centering
 \includegraphics[scale=0.42]{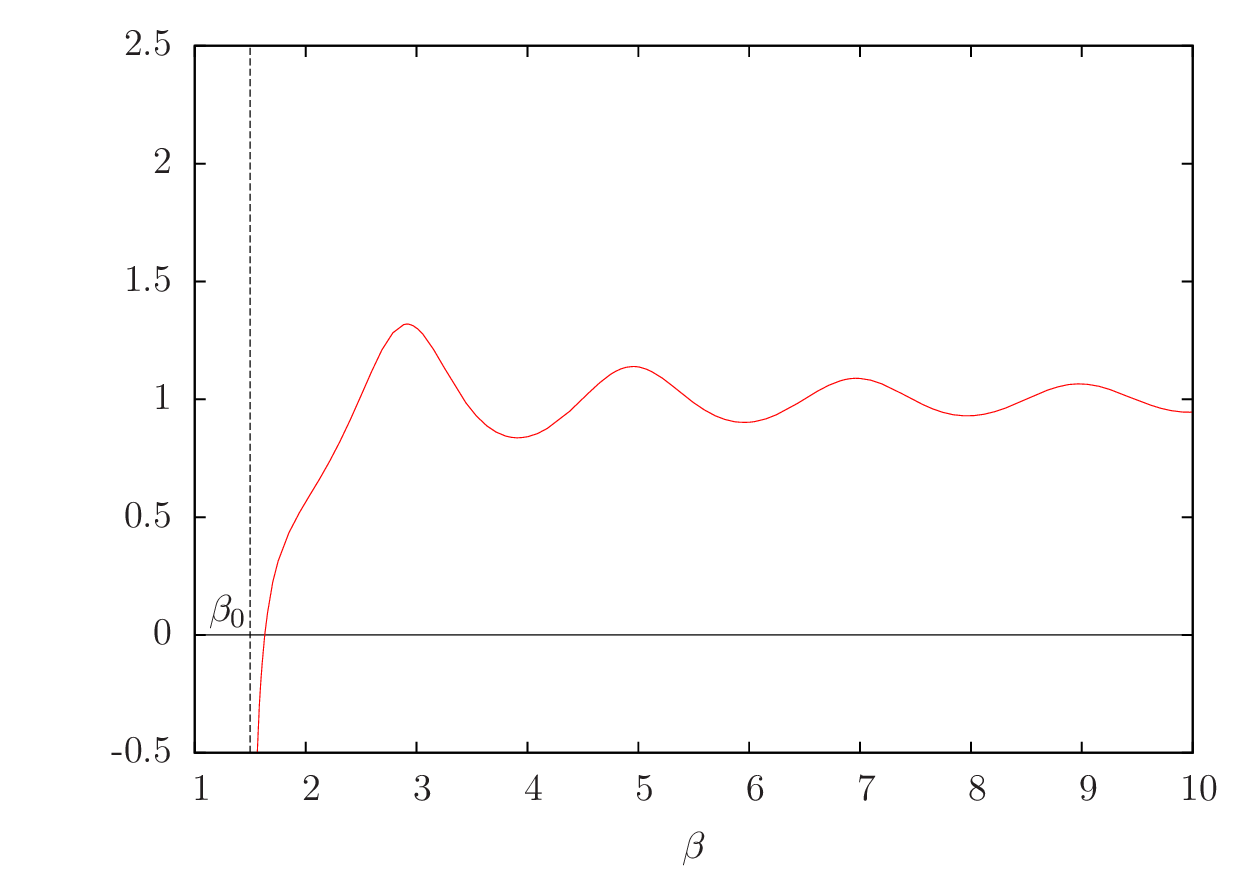}
 \includegraphics[scale=0.42]{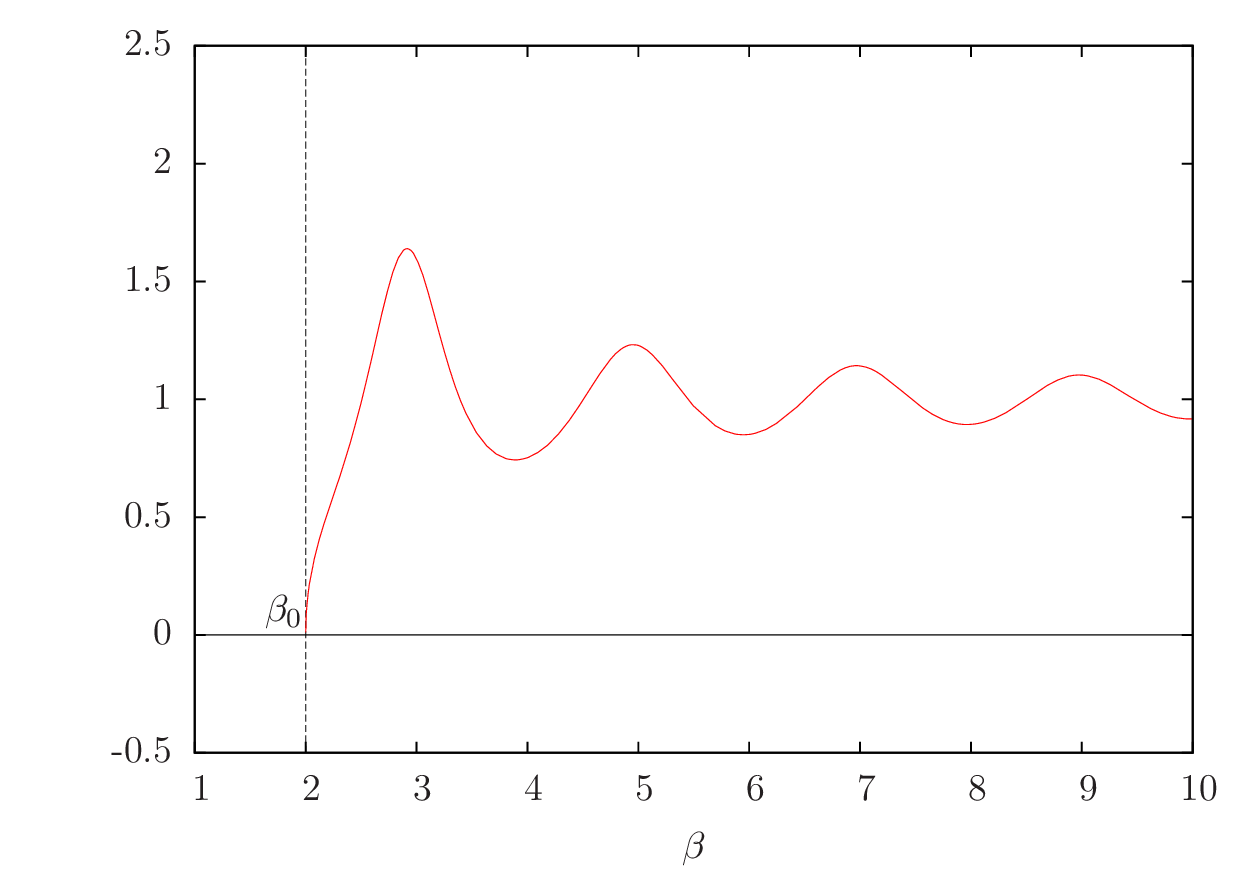}
 \includegraphics[scale=0.42]{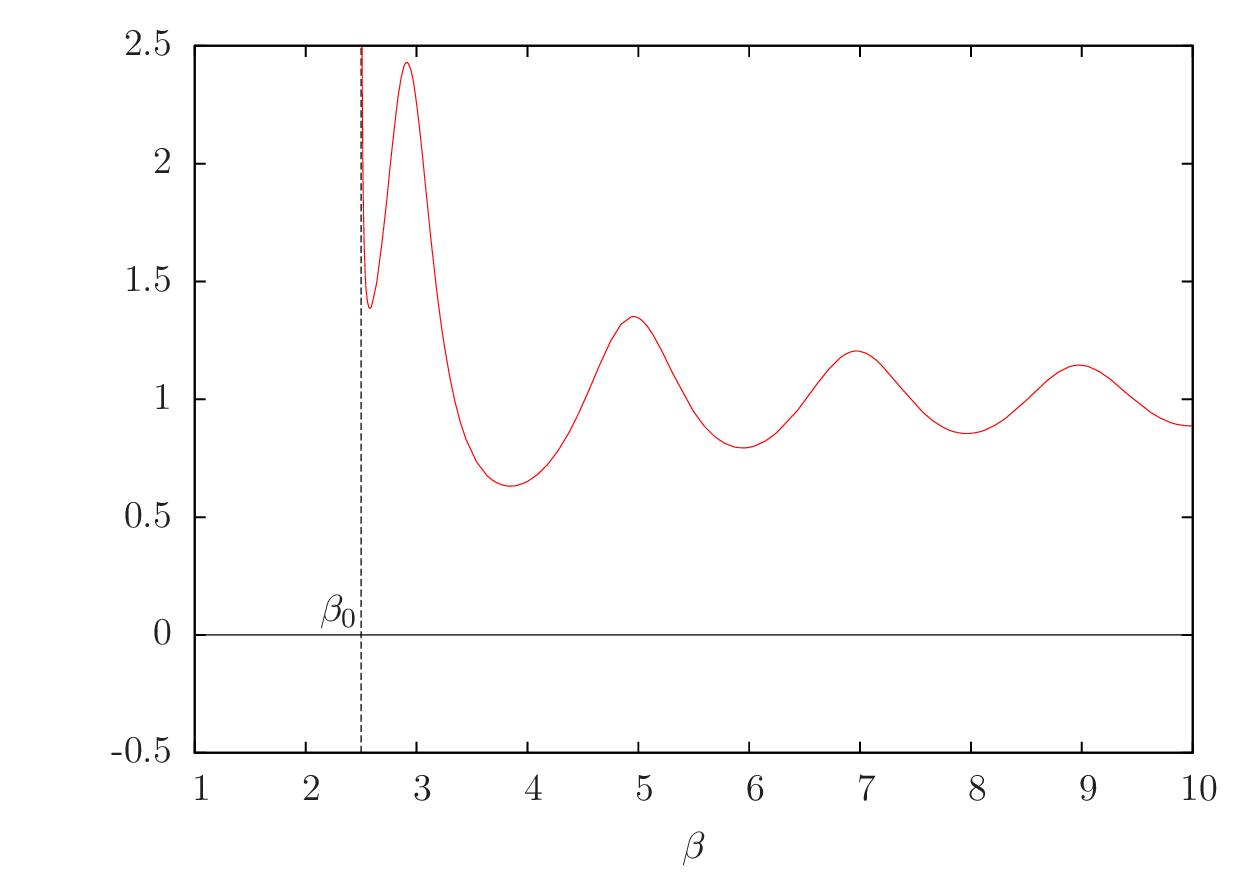}
 \includegraphics[scale=0.42]{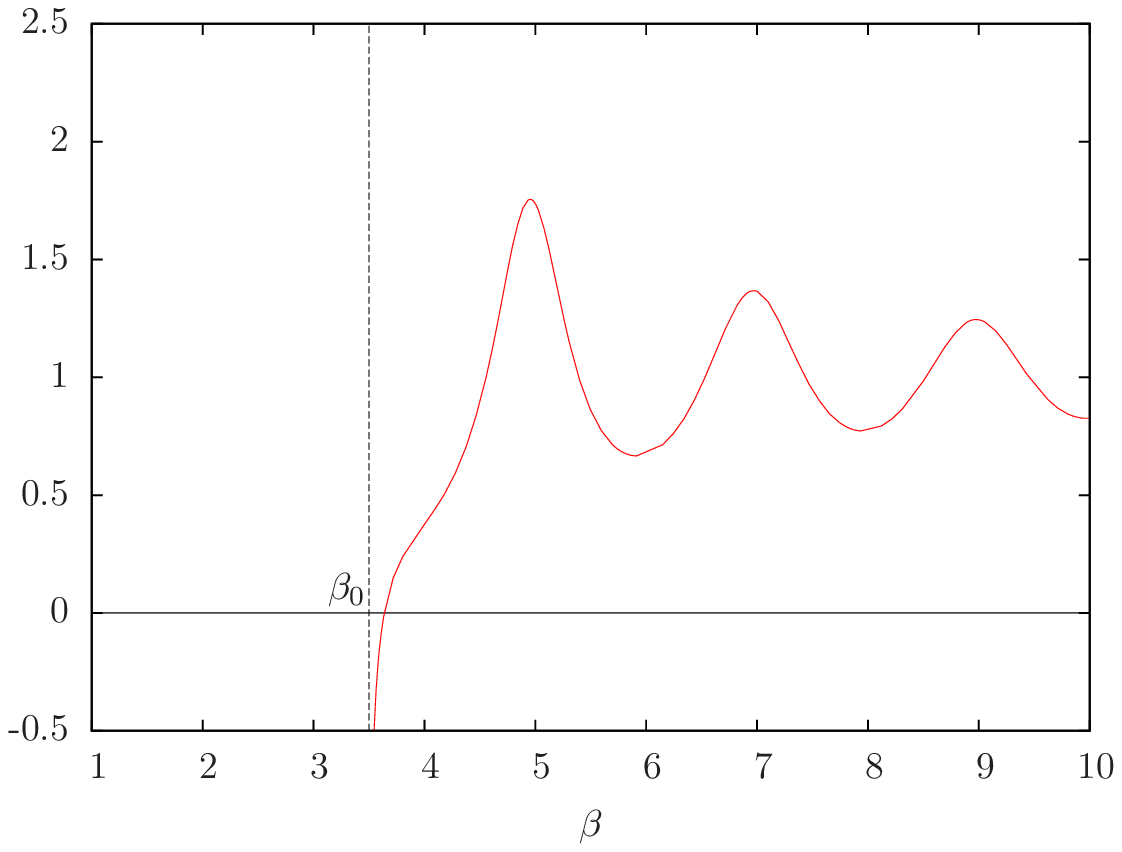}
 \includegraphics[scale=0.42]{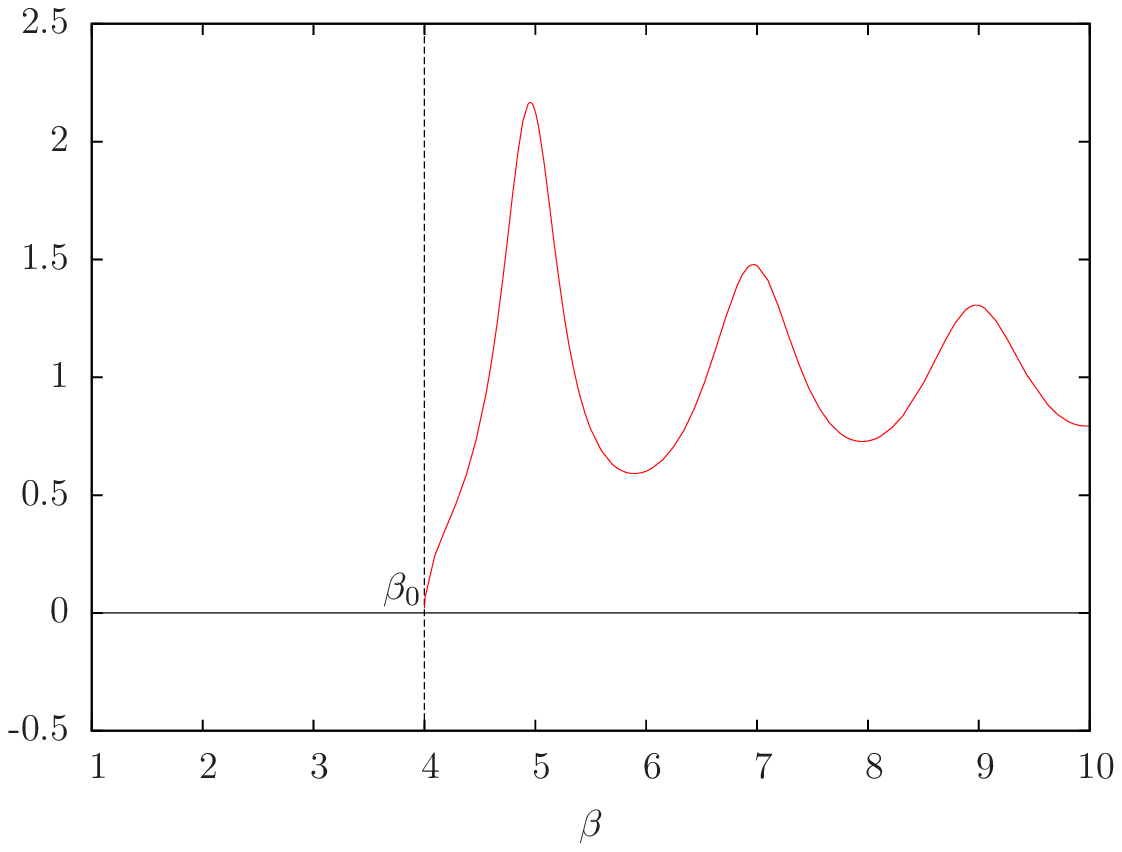}
 \includegraphics[scale=0.42]{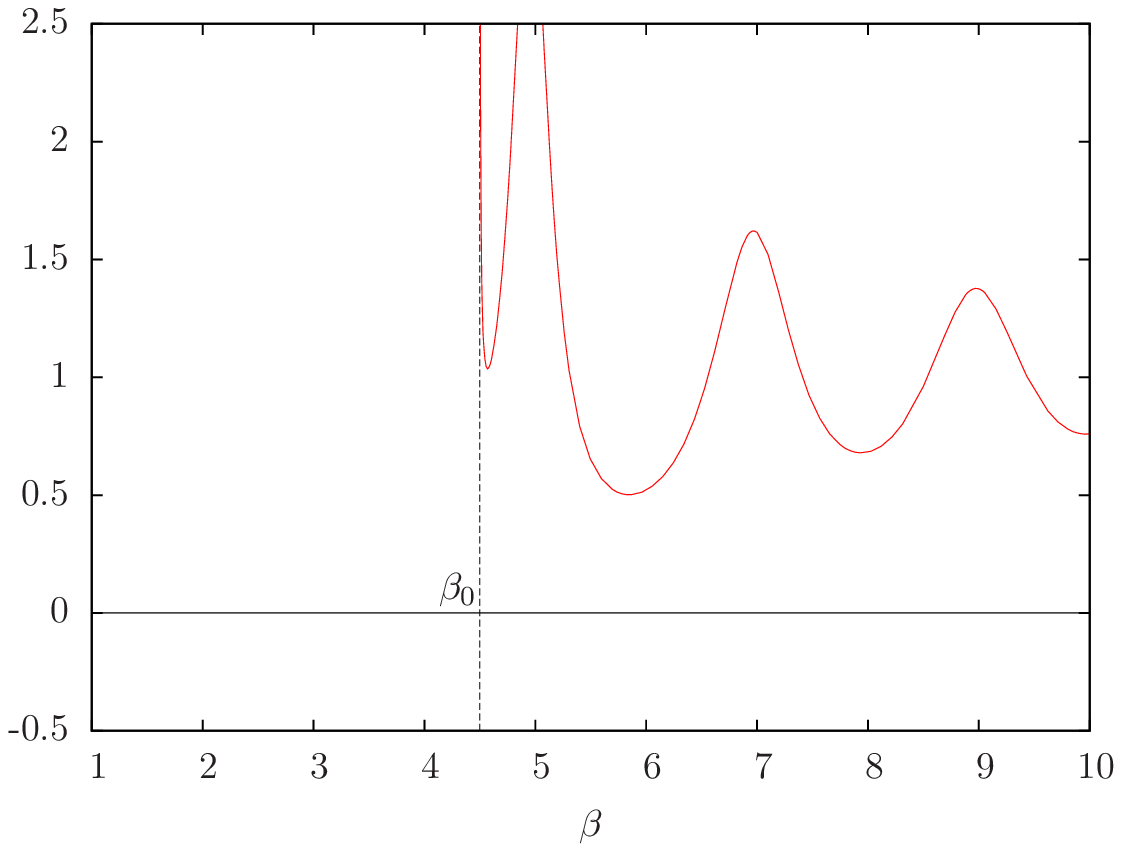}
\caption{Plots of the delay time $\tau$ (in units of $T/2$, where $T=\sqrt{\kappa/m}$ is the period associated to the harmonic oscillator) versus the ``energy'' $\beta$ of the incoming wave packet for six values of the step height $\beta_0$. (a) $\beta_0 = 1.5$, (b) $\beta_0 =2$, (c) $\beta_0= 2.5$, (d) $\beta_0 = 3.5$, (e) $\beta_0=4$, and (f) $\beta_0= 4.5$.}
\label{fig:delays}
\end{figure}

We prove that
\begin{gather}
\label{teo1} \lim_{\beta\rightarrow\infty}\delta^\prime(\beta) = \pi, \\
\label{teo2} \lim_{\beta\rightarrow\beta_0} \delta^\prime(\beta) = \begin{cases}
+\infty & \beta_0\in\bigcup_{k\in\mathbb{N}}\limits (2k,2k+1) \\
-\infty & \beta_0\in\bigcup_{k\in\mathbb{N}}\limits (2k+1,2k+2) \\
0 & \beta_0 \in\mathbb{N} = 1,2,\dots,
\end{cases}
\end{gather}
where the prime denotes the derivative with respect to $\beta$. To this purpose, note that by employing \eqreff{gammaformula} $\delta^\prime(\beta)$ can be cast in the form:
\begin{equation}\label{delayformula}
\delta^\prime\left(\beta\right) = \dfrac{\dfrac{1}{2}\sqrt{\beta-\beta_0}\sin\left(\beta\pi\right)\left[ \dfrac{1}{\beta-\beta_0}+\Psi\left(\dfrac{\beta}{2}\right)-\Psi\left(\dfrac{\beta+1}{2}\right)+\dfrac{2\pi}{\sin\left(\beta\pi\right)}\right]}{\dfrac{\left(\beta-\beta_0\right)\Gamma\left(\beta/2\right)}{\Gamma\left(\beta/2+1/2\right)\sqrt{2}}\sin^2\left(\dfrac{\beta\pi}{2}\right)+\dfrac{\Gamma\left(\beta/2+1/2\right)\sqrt{2}}{\Gamma\left(\beta/2\right)}\cos^2\left(\dfrac{\beta\pi}{2}\right)},
\end{equation}
where $\Psi$ is the Digamma function (that is, the logarithmic derivative of the Gamma function).\cite{AS1972} In \figname~\ref{fig:delays} we plot $\tau$ versus $\beta$ for different values of $\beta_0$. Note the resonances located at $\beta \simeq 3$, 5, 7, 9, $\dots$, corresponding to the formation of metastable states at the respective energies $E \simeq 5 \hbar\omega/2$, $9 \hbar\omega/2$, $13 \hbar\omega/2$, $17 \hbar\omega/2$, $\dots$. These states have lifetimes which decrease as the corresponding energies increase and move farther away from the threshold energy $U_0$. Conversely, as $U_0$ increases, the lifetime of the resonance closest to the height of the step becomes progressively longer and then infinite when the resonance turns into the next bound state. This behavior is evident in \figname~\ref{fig:delays}, in which the first three plots correspond to values of $\beta_0$ for which there is only one bound state. In the successive three plots the resonance at $\beta = 3$ has disappeared, having turned into the second bound state.

It is simple (using steepest descent or Stirling's formula, for example) to show that
\begin{equation}\label{eqn:j/2}
\dfrac{\Gamma(z+1/2)}{\Gamma(z)} = \sqrt{z}\left[1 + O\left(\dfrac{1}{z}\right)\right],
\end{equation}
for $z \gg 1$. By using one of the integral formulas for the Digamma function,\cite{AS1972} we can also show that
\begin{equation}\label{eqn:limdigamma}
\lim_{z\rightarrow\infty}\left[\Psi\left(z\right)-\Psi\left(z + \dfrac{1}{2}\right)\right] = 0.
\end{equation}
Thanks to \eqrefff{eqn:j/2}{eqn:limdigamma} it is straightforward to derive \eqrefff{teo1}{teo2}. In particular \eqreff{teo1} implies that
\begin{equation}
\lim_{\beta\rightarrow\infty}\tau(\beta) = \dfrac{\pi}{\omega} = \dfrac{T}{2}.
\end{equation}
The wave packet undergoes half an oscillation during the interaction with the harmonic potential before being reflected, which results in a delay of half a period compared with the reflection from a perfect mirror (that is, when the confining barrier is an infinite wall).
Thus, as expected, the high energy limit reproduces the classical behavior.

\section{Conclusions}\label{sec:Conclusions}

The main features of the discrete part of the spectrum can be summarized as follows. For sufficiently small $U_0$ (the height of the step) there is no discrete spectrum. When $U_0$ increases and approaches the value $\hbar\omega/2$ from below, there appears a resonance at energy $E_0 \simeq \hbar\omega/2$. This resonance converts into a bound state when $U_0$ reaches the value $\hbar \omega /2$ and the corresponding eigenfunction is proportional, at $x < 0$, to the eigenfunction of the ground state of the free harmonic oscillator and is flat otherwise. By further increasing the height of the step the ground state energy increases monotonically with $U_0$ and, as $U_0 \to \infty$, approaches asymptotically from below the first odd level $3 \hbar\omega/2$ of the full-space harmonic oscillator. A new discrete energy level appears at each energy $E_k = \hbar\omega(2k + 1/2)$ ($k \in \mathbb{N}$) whenever $U_0$ crosses the value $E_k$. In the limit of infinite $U_0$ (leading to the half-space oscillator), the energy levels become the odd levels of the oscillator itself, as expected from the symmetry of the problem. Loosely speaking, the levels are born as ``even'' and, upon increasing the height of the step, end up as ``odd'' (see \figname~\ref{energylevels}). This behavior is not peculiar to the step problem associated with the harmonic oscillator, but is typical of the corresponding step variant of every symmetric confining potential.

The continuous spectrum is simple and extends from $U_0$ to $\infty$. A wave packet coming from infinity collides with the confining harmonic branch and is thereby entirely reflected. The interaction with the potential results in a delay of the reflected packet which, as is well known for problems of this kind, is proportional to the derivative of the phase shift of the plane wave component evaluated at $\tilde{\beta} = \beta(\tilde{k})$. 
This delay can be interpreted as the interaction time with the harmonic barrier. When the confining part of the potential is infinite ($U(x) = +\infty$ at $x < 0$) the delay vanishes, and the reflection on a perfect mirror is instantaneous. The desirable feature of our example is that we can derive an exact analytic expression for the delay [\eqreff{delayformula}] as a function of the step height and of the peak energy of the incoming packet.

Although these characteristics are typical of the step variants of all symmetric confining potentials, the harmonic oscillator potential is ``more equal'' than the others. It is the only analytic (except possibly at $x=0$), convex or concave locally bounded symmetric and confining potential that gives rise to classical isochronous oscillations and thus to evenly spaced energy levels.\cite{Asorey:2007gb, Carinena:2007zz} In our step variant of the problem we recover both these features in the limit $U_0 \to \infty$, when the potential reduces to the half-space harmonic oscillator. Thus, it is possible that the harmonic potential is the only confining barrier that displays a constant nonvanishing interaction time in the limit of high energies. For steeper barriers we expect the interaction time $\tau$ to vanish at high energies, while for milder 
potentials we expect the delay to become infinite in this limit, in accordance with the corresponding classical behavior. Similarly, we expect that, as $U_0 \to \infty$, the spacing between two neighboring discrete levels approaches infinity in the former case and zero in the latter.

In a forthcoming paper we corroborate this conjecture by analyzing two examples using the integral representation method. This method can be employed to analyze a wide class of ``step-something'' potentials in which the harmonic part is replaced by another type of barrier. We encourage readers to investigate, for example, the step-linear (sl) and the step-exponential (se) potentials
\begin{align}
U_{\text{sl}}(x) & = \begin{cases}
U_0 & (x \geq 0) \\
-M x & (x < 0)
\end{cases}\\
U_{\text{se}}(x) &= \begin{cases}
U_0 & (x \geq 0) \\
Me^{-x/\sigma} & (x < 0)
\end{cases},
\end{align}
where $M$, $\sigma$, and $U_0$ are real positive constants.

\begin{acknowledgments}
We are grateful to Carlo Garoni for having inspired the topic that we have analyzed in this paper.
\end{acknowledgments}

\appendix

\section{Calculation of $J(\beta)$}\label{appendixA}
\begin{figure}
\centering
 \includegraphics[scale=0.7]{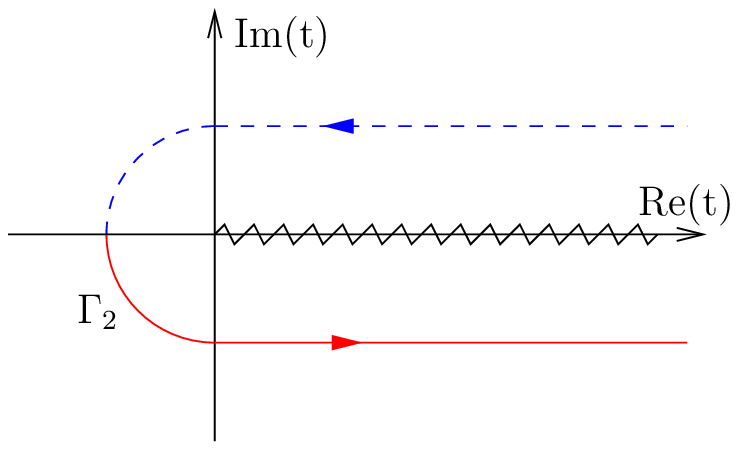}
 \includegraphics[scale=0.7]{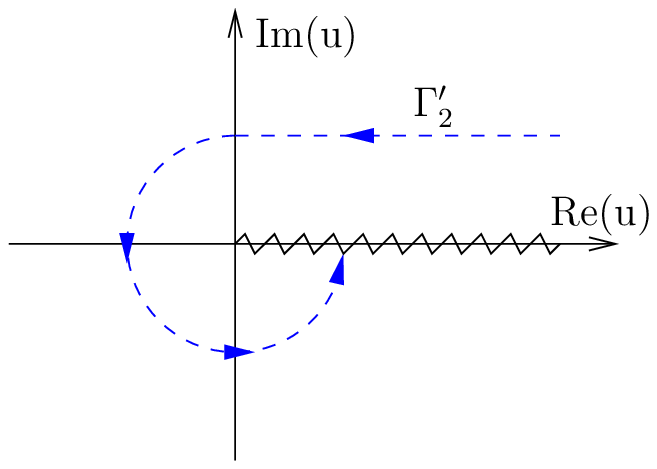}
 \includegraphics[scale=0.7]{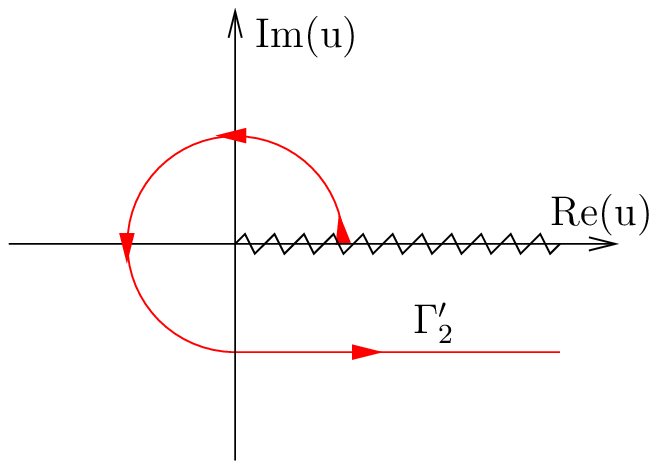}
\caption{The path $\Gamma_2$ and its transformed one $\Gamma_2^\prime$ after the change of variable $u=t^2$. When we take into account the presence of the cut, we must choose $0\leq \arg(u)<2\pi$ on the first sheet (dashed line), and $2\pi \leq \arg(u) <4\pi$ on the second one (solid line), that is, we choose the positive square root on the first sheet, and the negative one on the second.}\label{newpath}
\end{figure}

In this appendix we prove \eqreff{Jbeta_first}. According to the definition given in Sec.~\ref{sec:Eigenfunctions and Energy levels} we have
\begin{equation}\label{Jbeta}
J(\beta) \equiv F^{(2)}_\epsilon(0) = \!\int_{\Gamma_2}\ud t\, e^{-t^2} t^{-\beta},
\end{equation}
where $2\beta \equiv \epsilon + 1$. Assume $\beta\in\mathbb{C}$. For $\abs{\beta}\leq R$, the integrand function in \eqreff{Jbeta} is bi-continuous and holomorphic with respect to $\beta$ in any compact disc. Furthermore, its absolute value is bounded by a summable positive function: $
\abs{e^{-t^2} t^{-\beta}}\leq e^{-2\Re{(t^2)}}\abs{t}^{2R}$.
These properties imply that the integral in \eqreff{Jbeta} is uniformly convergent. Therefore, $J(\beta)$ is an entire function.
If we change the variable in \eqreff{Jbeta} to $u = t^2$, we must cut the complex plane and define $t = \sqrt{u}$ on its complete Riemann surface, which is composed of two sheets. The new path $\Gamma_2^\prime$ is shown in \figname~\ref{newpath}.

On both sheets of the $u$-plane, there is an integral along a straight line and an integral on a semi-loop. It is easy to show that the integral on the semi-loop vanishes when shrunk to a point, provided that $\beta\in(-\infty,1)$. The integrals along the straight lines can be deformed, using Cauchy's theorem, to become integrals along the positive real axis. We thus have
\begin{equation}
J(\beta) = -\dfrac{1}{2}\!\int_0^\infty\! \ud u\, e^{-u} u^{-\tfrac{\beta +1}{2}} + \dfrac{1}{2} e^{i4\pi\left(-\tfrac{\beta+1}{2}\right)}\!\int_0^\infty\!\ud u\, e^{-u} u^{-\tfrac{\beta +1}{2}},
\end{equation}
which we can write as
\begin{equation}\label{J}
J(\beta) = \dfrac{\sin(\pi\beta)}{ie^{i\pi\beta}}\Gamma\left(\dfrac{1-\beta}{2}\right).
\end{equation}
Because the poles of the Gamma function in \eqreff{J} are cancelled by the zeroes of the sine, the right-hand side is an entire function so that \eqreff{J} holds on the entire complex plane by analytic continuation.

\section{Taking limits under the integral sign}\label{appendixB}

For convenience we give here a very useful elementary theorem which we have used in the paper:

{\bf Theorem 1: The dominated convergence theorem}.
Let $f_k:\mathbb{R}\to\mathbb{R}$ be summable on an interval $I$, that is, $\int_I f_k < \infty$, $\forall k \in \mathbb{N}$. Moreover, let $f_k$ converge almost everywhere to a function $f_\infty:\mathbb{R}\to\mathbb{R}$. Suppose that there exists a positive $I$-summable function $g:\mathbb{R}\to\mathbb{R^{+}}$ that dominates every $f_k$ (that is, $\abs{f_k(x)}\leq g(x)\,\forall k\in\mathbb{N}$). It follows that $f_\infty$ is $I$-summable and that we can take the limit under the integral, that is,
\begin{equation}
\lim_{k\to\infty}\!\int_I\! \ud x\,f_k(x) = \!\int_I\!\ud x\,\lim_{k\to\infty}f_k(x) = \!\int_I\! \ud x\, f_\infty(x).
\end{equation}
It is noteworthy that, in our case, we do not need to invoke Lebesgue integration and the dominated convergence theorem, and our calculations are based on Riemannian integration. Even though Riemann integration theory lacks theorems regulating the interchange between limit and integration operations, the following theorem is sufficient for our purposes:\cite{C1967}

{\bf Theorem 2}.
Let $f_n$ be a sequence of functions defined on $[a,\infty)$ and Riemann integrable on $[a,b]$ for all $b>a$. Assume that (i)
$f_n(x)\to f_\infty(x)$ almost everywhere in $[a,\infty)$, $f_\infty$ being Riemann integrable on every finite interval. (ii) There exists a positive function $g$ defined on $[a,\infty)$ such that $\int_a^\infty g$ is convergent and $\abs{f_n(x)}\leq g(x)$ for all $n$.
Then $\int_a^\infty f_n \to \!\int_a^\infty f_\infty$.
Note that, in this theorem, the integrability of the limiting function is part of the hypothesis, whereas in the dominated convergence theorem it is a consequence of the theorem itself.


\begin{thebibliography}{6}

\bibitem{CF1976} A. Consortini and B. R. Frieden, ``Quantum-mechanical solutions for the simple harmonic oscillator in a box,'' Il Nuovo Cimento {\bf 35} (B2), 153--164 (1976).

\bibitem{ML1983} W. N. Mei and Y. C. Lee, ``Harmonic oscillator with potential barriers,'' J. Phys. A: Math. Gen. {\bf 16}, 1623--1632 (1983).

\bibitem{MC1988} J. L. Marin and S. A. Cruz, ``On the harmonic oscillator inside an infinite potential well,'' Am. J. Phys. {\bf 56}, 1134 (1988).

\bibitem{Barton:1990gp} G.~Barton, A.~J.~Bray, and A.~J.~McKane, ``The influence of distant boundaries on quantum mechanical energy levels,'' Am. J. Phys. {\bf 58}, 751--755 (1990).

\bibitem{GRD2005} V. G. Gueorguiev, A. R. P. Rau, and J. P. Draayer, ``Confined one-dimensional harmonic oscillator as a two-mode system,'' Am. J. Phys. {\bf 74} (5), 394--403 (2006).

\bibitem{MAS2007} H. E. Montgomery, Jr., N. A. Aquino, and K. D. Sen, ``Degeneracy of confined D-dimensional
harmonic oscillator,'' Int. J. Quantum Chemistry {\bf 107}, 798--806 (2007).

\bibitem{Robinett2000A} R. W. Robinett, ``Visualizing the collapse and revival of wave packets in the infinite square well using expectation values,'' Am. J. Phys. {\bf 68} (5), 410--420 (2000).

\bibitem{Robinett2000B} R. W. Robinett, ``Wave packet revivals and quasirevivals in one-dimensional power law potentials,'' J. Math. Phys. {\bf 41}, 1801--1813 (2000).

\bibitem{Chalk1990} J. D. Chalk, ``Tunneling through a truncated harmonic oscillator potential barrier,'' Am. J. Phys. {\bf 58} (2), 147--151 (1990).

\bibitem{Prosperi} P. Caldirola, R. Cirelli, and G. M. Prosperi, \textsl{Introduction to Theoretical Physics} (UTET, Turin 1982), in Italian.

\bibitem{griffiths} D. J. Griffiths, \textsl{Introduction to Quantum Mechanics}, 2nd. ed. (Benjamin Cummings, San Francisco 2004)

\bibitem{BC:1980} M. Bowen and J. Coster, ``Methods of establishing the asymptotic behavior of the harmonic oscillator wave functions,'' Am. J. Phys. {\bf 48}, 307--308 (1980).

\bibitem{AS1972} M. Abramowitz and I. A. Stegun, \textsl{Handbook of Mathematical Functions}, 10th. ed. (Dover Publications, New York, 1972)

\bibitem{Hochstadt1976} H. Hochstadt, \textsl{The Functions of Mathematical Physics} (Dover Publications, New York, 1976), pp. 100--105.

\bibitem{C1967} F. Cunningham, Jr., ``Taking limits under the integral sign,'' Mathematics Magazine {\bf 40} (4), 179--186 (1967).

\bibitem{foot1}Because $\abs{y}$ is a positive real quantity, there are no complications with the multi-valued function $(y + z)^\beta$. Those would instead arise if we were to extract $y^\beta$ for negative values of $y$.

\bibitem{Carinena:2007zz}J.~F.~Cari\~nena, A.~M.~Perelomov, and M.~F.~Ra\~nada, ``Isochronous classical systems and quantum systems with equally spaced spectra,'' J. Phys. Conf. Ser. {\bf 87}, 012007 (2007).

\bibitem{Asorey:2007gb}M.~Asorey, J.~F.~Cari\~nena, G.~Marmo, and A.~Perelomov, ``Isoperiodic classical systems and their quantum counterparts,'' Ann. Phys. (NY) {\bf 322}, 1444--1465 (2007).

\end{thebibliography}
\end{document}